\documentclass[a4paper,11pt]{article}
\usepackage{pos}
\usepackage{cleveref}
\usepackage{floatrow}
\usepackage{amsmath}
\usepackage{xspace}

\newcommand{\gev}{\;\text{GeV}\xspace}

\title{New constraints on extended scalar sectors from the trilinear Higgs coupling}

\author[a,b]{Henning Bahl}
\author*[c]{Johannes Braathen}
\author[c,d]{Georg Weiglein}

\affiliation[a]{University of Chicago, Department of Physics, 5720 South Ellis Avenue, Chicago, IL~60637~USA}
\affiliation[b]{Institut für Theoretische Physik, Philosophenweg 16, 69120 Heidelberg, Germany}
\affiliation[c]{Deutsches Elektronen-Synchrotron DESY, Notkestr. 85, 22607 Hamburg, Germany}
\affiliation[d]{II. Institut für Theoretische Physik, Universität Hamburg, Luruper Chaussee 149, 22761 Hamburg, Germany}

\emailAdd{bahl@thphys.uni-heidelberg.de}
\emailAdd{johannes.braathen@desy.de}
\emailAdd{georg.weiglein@desy.de}

\abstract{
The trilinear Higgs coupling $\lambda_{hhh}$ is a crucial tool to probe the structure of the Higgs potential and to search for possible effects of physics beyond the Standard Model (SM). Focusing on the Two-Higgs-Doublet Model as a concrete example, we identify parameter regions in which $\lambda_{hhh}$ is significantly enhanced with respect to its SM prediction. Taking into account all relevant corrections up to the two-loop level, we show that current experimental bounds on $\lambda_{hhh}$ already rule out significant parts of the otherwise unconstrained parameter space. We illustrate the interpretation of the current results and future measurement prospects on $\lambda_{hhh}$ for a benchmark scenario. Recent results from direct searches for BSM scalars in the $A\to ZH$ channel and their implications will also be discussed in this context.
}

\FullConference{The European Physical Society Conference on High Energy Physics (EPS-HEP2023)\\
 21-25 August 2023\\
Hamburg, Germany\\}

\begin{document}

\begin{flushright}
DESY-23-163
\end{flushright}

\maketitle

\section{Introduction\vspace{-0.3cm}}
The trilinear Higgs coupling $\lambda_{hhh}$ determines the shape of the Higgs potential and is a crucial probe of the dynamics of the electroweak phase transition and of phenomena beyond the Standard Model (BSM). Sizeable deviations from the Standard Model (SM) value are known to occur commonly in many BSM theories~\cite{Kanemura:2004mg}. A prime task of the LHC and of future colliders is thus to measure $\lambda_{hhh}$ as precisely as possible, in particular through the process of (non-resonant) Higgs-pair production. At present, the best bounds on $\lambda_{hhh}$, expressed in terms of the coupling modifier $\kappa_\lambda\equiv \lambda_{hhh}/(\lambda_{hhh}^\text{SM})^{(0)}$, are $-0.4<\kappa_\lambda<6.3$ at 95\%~C.L.~\cite{ATLAS:2022jtk}, while more precise determinations are expected at future colliders~\cite{deBlas:2019rxi}.

In Ref.~\cite{Bahl:2022jnx}, we demonstrated that the current experimental information on $\kappa_\lambda$ already puts severe constraints on otherwise unconstrained parameter regions of BSM models with extended Higgs sectors, having taken a Two-Higgs-Doublet Model (2HDM) as a concrete example. This follows from the observation that radiative corrections can enhance $\kappa_\lambda$ by up to a factor 10. In this context, we showed the importance of incorporating numerically significant two-loop corrections to $\lambda_{hhh}$, first computed in Refs.~\cite{hhh2L}. The origin of these large corrections can be traced back to large, but \emph{still perturbative}, trilinear and quartic couplings between the SM-like and the BSM Higgs bosons, and importantly these can appear not only in the 2HDM but in many models with extended Higgs sectors. As a result for the 2HDM, significant parts of the parameter space that are not fine-tuned and unconstrained so far are now excluded.
In these proceedings, we summarise our results of Ref.~\cite{Bahl:2022jnx}, and update them in light of new experimental results, in particular Ref.~\cite{ATLAS:2023szc}.

\vspace{-0.2cm}
\section{Constraints on BSM parameter space arising from $\lambda_{hhh}$\vspace{-0.3cm}}
We consider here a CP-conserving 2HDM and we refer the reader to Ref.~\cite{Bahl:2022jnx} for a detailed description of the model and our conventions (see also Ref.~\cite{Kanemura:2004mg}). Throughout this work, we identify the lightest CP-even mass eigenstate, $h$, with the observed 125-GeV Higgs boson and take the alignment limit, $i.e.$ $\alpha = \beta - \pi/2$~\cite{Gunion:2002zf}. 
This ensures that the tree-level couplings of $h$ are exactly equal to their SM values, and in particular that $(\lambda_{hhh})^{(0)}=(\lambda_{hhh}^\text{SM})^{(0)} = 3m_h^2/v$. The remaining input parameters for the numerical analysis are $m_{H}$, $m_{A}$, $m_{H^\pm}$, $M^{2} = m_{12}^2/(\sin\beta\cos\beta)$, and $\tan\beta\equiv v_2/v_1$. For our one- (1L) and two-loop (2L) predictions, we employ results for $\lambda_{hhh}$ from Refs.~\cite{hhh2L} including the dominant 2L corrections (arising from heavy BSM scalars as well as the top quark) in a number of BSM models, including an aligned version of the 2HDM. The largest type of quartic coupling entering corrections to $\lambda_{hhh}$, both at one and two loops, are those between two SM-like and two heavy BSM Higgs bosons, of the form $g_{hh\Phi\Phi}=-2(M^2-m_\Phi^2)/v^2$, where $\Phi =H, A, H^\pm$ and where one external Higgs boson $h$ is potentially replaced by a vacuum expectation value. 

The experimental limits on $\kappa_\lambda$ from Higgs-boson pair production~\cite{ATLAS:2022jtk} do not only rely on the assumption that all other Higgs couplings are SM-like (which is true in the alignment limit) but also that non-resonant Higgs-boson pair production only deviates from the SM via a modified value of $\lambda_{hhh}$. However, in the 2HDM, the additional Higgs bosons can in principle cause further modifications of Higgs-boson pair production. While resonant contributions with an $H$ or an $A$ boson in the $s$ channel are zero in our case due to alignment and CP conservation, at the loop level the BSM Higgs bosons can also contribute beyond their effects on $\lambda_{hhh}$ -- $e.g.$ via a correction to the box diagram. Nevertheless, our calculation includes the leading corrections to Higgs-boson pair production in powers of $g_{hh\Phi\Phi}$ (at NLO and NNLO), which we find to be the source of the large loop corrections in our numerical investigation. We therefore expect our calculation to capture the dominant effects on Higgs-boson pair production, justifying the application of the experimental limit on $\kappa_\lambda$ for the comparison with our predictions. 

\vspace{-0.2cm}
\section{Numerical results\vspace{-0.3cm}}
\paragraph*{Parameter scan:}
We begin our numerical study by performing a parameter scan of the 2HDM parameter space, in order to identify regions with significant BSM enhancements in $\kappa_\lambda$. For concreteness, we consider here a 2HDM of type I, although it should be emphasised that similar results are expected for all 2HDM types.\footnote{The difference between the 2HDM types appears only in the down-type and lepton Yukawa couplings, which do not enter the dominant corrections to $\lambda_{hhh}$ employed in our work.} The theoretical and experimental constraints that we include in our scan are the following: \textit{(i)} vacuum stability and boundedness-from-below (BFB) of the Higgs potential; \textit{(ii)} NLO perturbative unitarity~\cite{nlopertunit}; \textit{(iii)} electroweak precision observables (EWPO) calculated at 2L with \texttt{THDM\_EWPOS}~\cite{THDMEWPOS}; \textit{(iv)} compatibility of the SM-like scalar with the experimentally discovered Higgs boson using \texttt{HiggsSignals}~\cite{Bechtle:2020uwn}; \textit{(v)} limits from direct searches for BSM scalars using \texttt{HiggsBounds}~\cite{HiggsBounds}; \textit{(vi)} $b$ physics~\cite{Haller:2018nnx}. All these constraints are applied using \texttt{ScannerS}~\cite{Muhlleitner:2020wwk} -- except \textit{(ii)} and \textit{(iii)}, which we evaluate separately. For parameter points passing all constraints, we compute $\kappa_\lambda$ at one and two loops --- which we denote $\kappa_\lambda^{(1)}$ and $\kappa_\lambda^{(2)}$. For the parameter scan, we impose $M_h=125\text{ GeV}$ and $\alpha = \beta - \pi/2$, while the BSM scalar masses are varied in the range $[300\text{ GeV}, 1500\text{ GeV}]$, $\tan\beta$ in $[0.8,50]$, and $m_{12}^2$ in $[0,\ 4\cdot 10^6\text{ GeV}^2]$.

\begin{figure}
    \centering    
    \includegraphics[width=.95\textwidth]{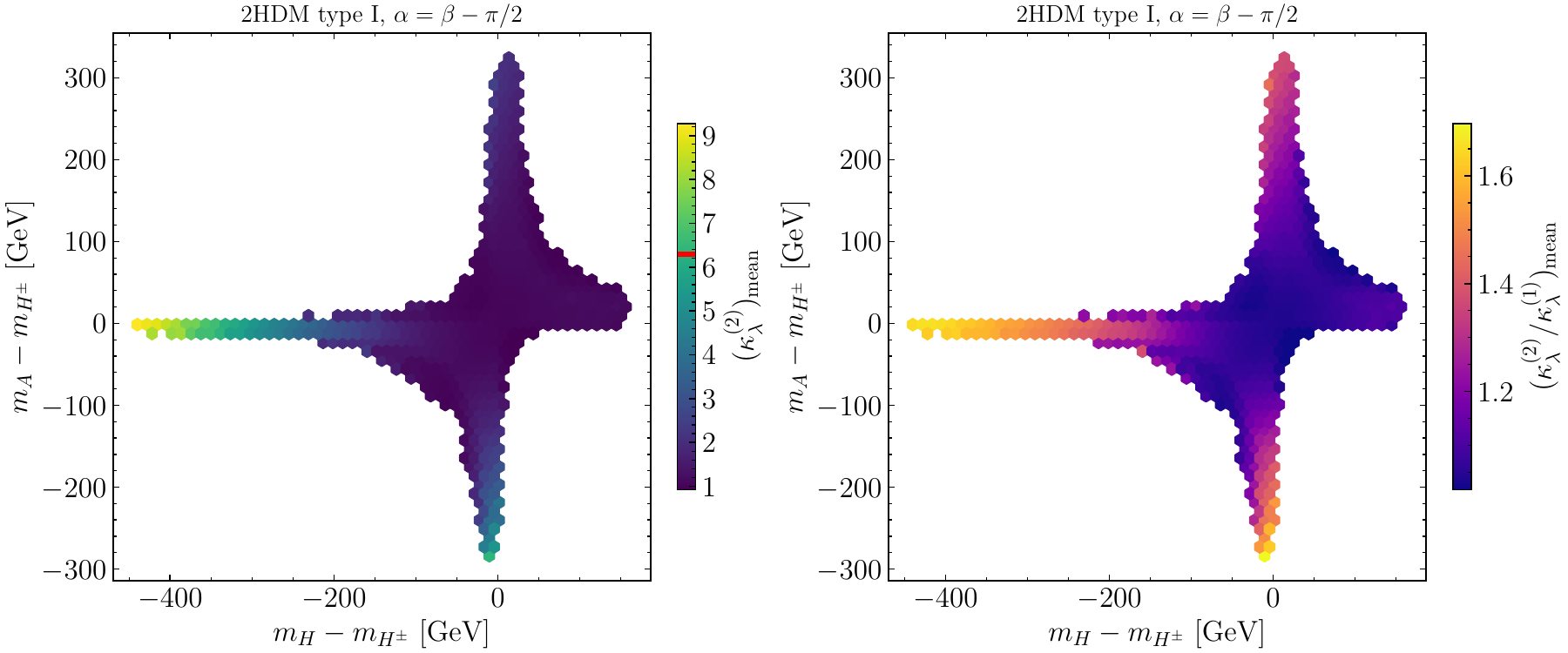}
    \caption{Results of a parameter scan in the 2HDM of type-I, shown in the $(m_H - m_{H^\pm}, m_A - m_{H^\pm})$ plane. \textit{Left:} the colour indicates the mean of $\kappa^{(2)}_\lambda$ in each hexagon-shaped patch; \textit{right:} the colour indicates the mean of $\kappa^{(2)}_\lambda/\kappa^{(1)}_\lambda$ in each patch. In the colour bar of the left-hand plot, the red line corresponds the current experimental upper limit on $\kappa_\lambda$~\cite{ATLAS:2022jtk}. \vspace{-0.5cm}}
    \label{fig:scan}
\end{figure}

Fig.~\ref{fig:scan} presents results of our parameter scan in the plane $(m_H - m_{H^\pm}, m_A - m_{H^\pm})$, with all shown points passing the constraints \textit{(i)}-\textit{(vi)} described above. In the left panel of Fig.~\ref{fig:scan}, we display for every hexagon-shaped patch the mean value (computed over all scan points within a patch) of the prediction of $\kappa_\lambda$ including corrections up to 2L, which we denote $\kappa_\lambda^{(2)}$. The ``cross-like'' shape of the yet-unconstrained region is caused by the EWPO, which enforce either $m_{H} \simeq m_{H^\pm}$ or $m_{A} \simeq m_{H^\pm}$, while the boundaries of the regions are determined by perturbative unitarity and BFB. The largest corrections to $\lambda_{hhh}$ are found for $m_A\simeq m_{H^\pm}$ and $m_H-m_{H^\pm}\lesssim -300\ \text{GeV}$, and to a lesser extent for $m_H\simeq m_{H^\pm}$ and $m_A-m_{H^\pm}\lesssim -220\ \text{GeV}$. In particular, for $m_{A}\simeq m_{H^\pm}$ and $m_{H}-m_{H^\pm}\lesssim -375\ \text{GeV}$, $\kappa_\lambda$ reaches maximal values of nearly $\sim 10$, well beyond the current experimental limit of $6.3$ --- indicated by the red line in the colour bar. This means that the present experimental limits on $\kappa_\lambda$ already have a strong impact on the viable 2HDM parameter space. The largest deviations in $\kappa_\lambda$ are found for $m_{H} \simeq M < m_{A} \simeq m_{H^\pm}$. This is explained through the interplay between the size of the different $g_{hh\Phi\Phi}$ couplings, which enter the corrections to $\lambda_{hhh}$ and increase with the difference between the BSM mass scale $M$ and the BSM scalar masses, and the constraints from perturbative unitarity and BFB. These two constraints are more stringent in the regions where $m_{A} < m_{H} \simeq m_{H^\pm}$ or where $m_{H} > m_{A} \simeq m_{H^\pm}$ than in the one where $m_{H} < m_{A} \simeq m_{H^\pm}$. This translates into smaller allowed splittings between $M$ and the BSM scalar masses, and hence into smaller quartic couplings in the former regions. Next, in the right panel of Fig.~\ref{fig:scan}, we show for each hexagon-shaped patch the mean value of $\kappa_\lambda^{(2)}/\kappa_\lambda^{(1)}$. The largest (in relative size) two-loop corrections occur for $m_{H} < m_{A} \simeq m_{H^\pm}$, and to a lesser extent for $m_{A} < m_{H} \simeq m_{H^\pm}$ and $m_{H} \simeq m_{H^\pm} < m_{A}$. Fig.~\ref{fig:scan} illustrates that the parameter region where the mean value of $\kappa_\lambda^{(2)}$ is largest coincides with that where the 2L corrections are most important --- reaching almost 70\% of the 1L contributions. Taking into account relevant 2L corrections is therefore crucial to reliably compare predictions for $\lambda_{hhh}$ with the corresponding experimental bounds. We note finally that the large 2L corrections encountered here do not indicate a breakdown of perturbation theory: indeed, all parameter points shown here pass the criterion of NLO perturbative unitarity. Furthermore, we estimated by dimensional analysis the size of the corresponding dominant three-loop (3L) corrections, and we found for all allowed points that the 3L contributions are significantly smaller than the 2L ones.

\begin{figure}
\floatbox[{\capbeside\thisfloatsetup{capbesideposition={right,center},capbesidewidth=4cm}}]{figure}[\FBwidth]
{\caption{Benchmark plane of the 2HDM of type I, varying the two scales $M=m_H$ and $m_A=m_{H^\pm}$, and for $\tan\beta=2$ and $\alpha=\beta-\pi/2$. The star indicates the location of a $2.9\sigma$ excess at about $m_H=450$ GeV and $m_A=650$ GeV found in Ref.~\cite{ATLAS:2023szc}. The coloured lines and regions are explained in the legend and in the text. }\label{fig:benchmark}}
{ \includegraphics[width=.5\textwidth]{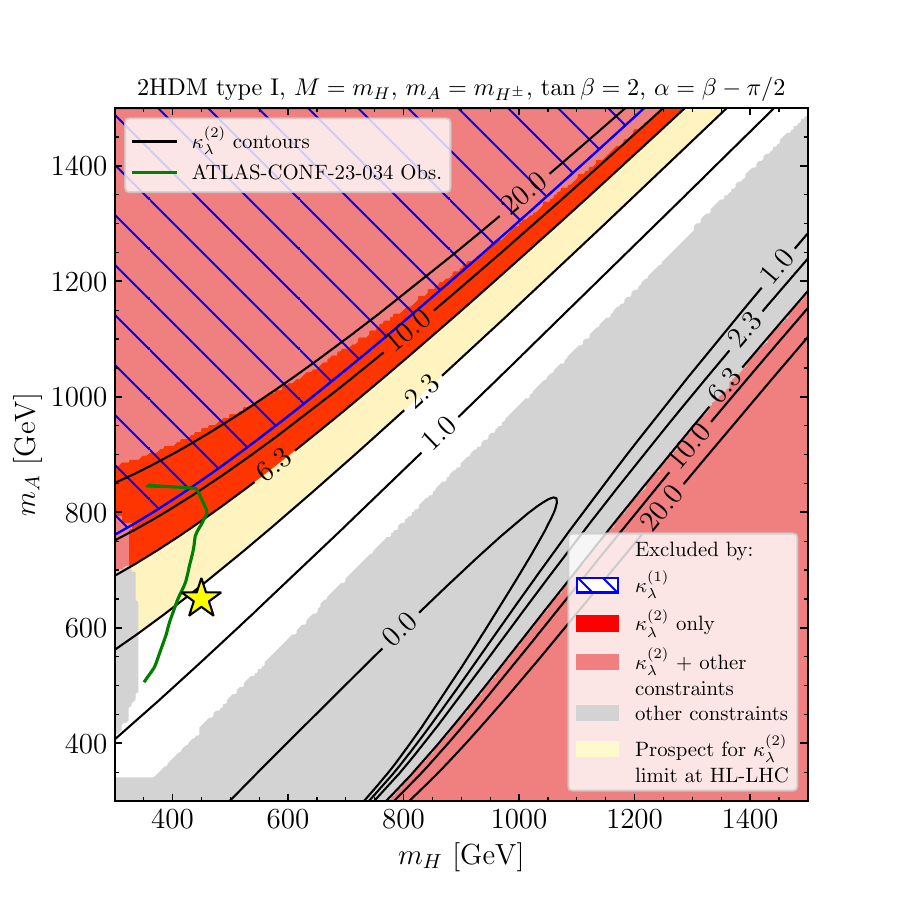}} 
\vspace{-0.5cm}
\end{figure}

\vspace{-0.2cm}
\paragraph*{Benchmark scenario:}
To illustrate the impact of present and future experimental information on $\kappa_\lambda$ on the 2HDM parameter space, and inspired by the previous parameter scan, we consider a benchmark scenario with two BSM mass scales (that will be varied) $M = m_{H}$, $m_{A} = m_{H^\pm}$, and where $\tan\beta=2$ and $\alpha = \beta - \pi/2$. 
We present in Fig.~\ref{fig:benchmark} the resulting $(m_H,m_A)$ parameter plane. The coloured areas indicate parameter regions that are excluded by one or more of the various constraints. The region that is excluded on the basis of constraints \textit{(i)}-\textit{(vi)} listed above is displayed in grey, while the region that is excluded both by these other constraints and the current constraint on $\kappa_\lambda$ is shown in light red. The dark red area indicates the parameter region that is only excluded by the current constraint on the trilinear Higgs coupling, taking into account 2L corrections in its calculation. Contour lines for constant values of $\kappa_\lambda^{(2)}$ are shown in black.

The existing constraints on $\lambda_{hhh}$ exclude large parts of the benchmark plane, namely in the upper left and lower right parts of the plane. Whereas the lower right part of the plot is also excluded by other constraints (mainly by BFB, but also partly by perturbative unitarity), the excluded region in the upper left part of the plot significantly exceeds the one that is covered also by other constraints (mainly by perturbative unitarity). We find that the part of the parameter plane stretching from around $(m_H, m_A)\simeq (300, 800)\gev$ to $(m_H, m_A)\simeq (1250, 1500)\gev$ is only excluded by the constraints on the trilinear Higgs coupling, incorporating 2L corrections into its prediction. If instead only 1L corrections had been taken into account for the prediction of $\kappa_\lambda$, the impact of the constraint on $\lambda_{hhh}$ would appear much smaller (blue hatched area). Meanwhile, the yellow region illustrates the additional region that will be probed by the constraint from $\kappa_\lambda$ at the HL-LHC, where the expected 95\% bounds are $0.1<\kappa_\lambda<2.3$  (with $3\text{ ab}^{-1}$ data~\cite{Cepeda:2019klc} and assuming SM rates). 

Finally, we discuss for this benchmark plane the impact of the recent results~\cite{ATLAS:2023szc} from direct searches for BSM scalars in the $A\to Z H$ channel (see also Ref~\cite{Biekotter:2023eil}). The solid green line corresponds to the observed search limit, while the star, located at $m_A=650\text{ GeV}$ and $m_H=450\text{ GeV}$, indicates an observed excess with a (local) significance of $2.9\sigma$. Interestingly, this excess is found very close to the yellow shaded region probed with $\kappa_\lambda$ at the HL-LHC, illustrating the important interplay between direct and indirect probes of the 2HDM parameter space at colliders.


\vspace{-0.5cm}
\section{Summary\vspace{-0.3cm}}

A precise determination of the trilinear Higgs coupling is essential to gain access to the shape of the Higgs potential and to probe possible effects of BSM physics. We showed that confronting the lates{}t experimental bounds on $\lambda_{hhh}$ with theoretical predictions incorporating numerically important 2L contributions allows excluding significant parts of the BSM parameter space that would otherwise remain unconstrained. These results have important implications for searches at the LHC and future colliders and show the crucial role played by $\lambda_{hhh}$ to discriminate between different possible manifestations of the underlying physics of e{}lectroweak symmetry breaking. Considering as a concrete example an aligned 2HDM and taking into account other rel{}evant theoretical and ex{}perimental constraints, we found that large BSM radiative corrections can enhance $\lambda_{hhh}$ by up to an order of magnitude compared to its SM value. In this context, it is particularly important to incorporate the dominant 2L corrections, which can reach about 70\% of the 1L contribution. We investigated a benchmark scenario exhibiting large BSM deviations in the trilinear Higgs coupling and we discussed the impact of present and expected future bounds on $\lambda_{hhh}$, as well as the interplay with direct searches for additional Higgs bosons. Our analysis places new exclusion bounds on parameter regions that up to now were in agreement with all relevant constraints. Finally, we emphasise that our findings, discussed here for a 2HDM, apply more generally for a wide range of BSM models; this was also shown recently in Ref.~\cite{Bahl:2023eau} with the public tool \texttt{anyH3}, which allows full 1L calculations of $\lambda_{hhh}$ for arbitrary renormalisable models. 

\vspace{-0.2cm}
\paragraph*{Acknowledgements}
\sloppy{J.B.\ and G.W.\ acknowledge support by the Deutsche Forschungsgemeinschaft (DFG, German Research Foundation) under Germany‘s Excellence Strategy --- EXC 2121 ``Quantum Universe'' --- 390833306. H.B.\ acknowledges support by the Alexander von Humboldt foundation. J.B.\ is supported by the DFG Emmy Noether Grant No.\ BR 6995/1-1. This work has been partially funded by the Deutsche Forschungsgemeinschaft 
(DFG, German Research Foundation) --- 491245950. }

\end{document}